\documentclass[english,aps,floatfix]{revtex4}
\usepackage[T1]{fontenc}
\usepackage[latin9]{inputenc}
\usepackage{graphicx}
\usepackage{dsfont}

\usepackage{slashed}

\usepackage{amsfonts}

\usepackage[dvips]{epsfig}

\usepackage{babel}

\usepackage{babel}

\usepackage{babel}

\usepackage{babel}

\usepackage{babel}

\usepackage{babel}

\begin{document}

\title{A family of loss-tolerant quantum coin flipping protocols}

\author{N. Aharon\textsuperscript{1}, S. Massar\textsuperscript{2}
and J. Silman\textsuperscript{2}}

\affiliation{\textsuperscript{1}School of Physics and Astronomy, Tel-Aviv University,
Tel-Aviv 69978, Israel \\
 \textsuperscript{2}Laboratoire d'Information Quantique, Universit\'{e}
Libre de Bruxelles, 1050 Bruxelles, Belgium}
\begin{abstract}
\textbf{We present a family of loss-tolerant quantum strong coin flipping
protocols; each protocol differing in the number of qubits employed.
For a single qubit we obtain a bias of $\mathbf{0.4}$, reproducing
the result of} \textbf{Berl\'{i}n} \textbf{\emph{et al}}\textbf{.} \textbf{{[}}Phys.
Rev. A \textbf{80}, 062321 (2009)\textbf{{]}, while for two qubits
we obtain a bias of $\textbf{0.3975}$. Numerical evidence based on
semi-definite programming indicates that the bias continues to decrease
as the number of qubits is increased but at a rapidly decreasing rate.}
\end{abstract}
\maketitle

\section{Introduction}

Coin flipping (CF) is a cryptographic task in which a pair of remote
distrustful parties, usually referred to as Alice and Bob, must agree
on a random bit. The problem was first introduced in 1981 by Blum
\cite{Blum}, who studied it in classical settings. There are two
variants of the problem: `strong' CF (SCF) and `weak' CF (WCF). In
SCF each party is not aware of the other's preference for the coin's outcome.
In contrast, in WCF the parties have opposite and
known preferences. Hence, in WCF there is always a winner and a loser,
unless one of the parties is caught cheating, in which case the protocol
is aborted. Let $P_{xy}$ denote the probability that the Alice (Bob)
obtains the outcome $x$ ($y$) and let $P_{\perp}$ denote the probability
that the protocols is aborted. If both parties are honest then $P_{00}=P_{11}=1/2$
and $P_{10}=P_{01}=P_{\perp}=0$, i.e. the parties always agree on
the outcome of the coin. The security of a CF protocol is quantified
by the extent to which dishonest parties can bias the outcome. We
denote by $P_{*i}\hat{=}1/2+\epsilon_{*i}$ ($P_{i*}\hat{=}1/2+\epsilon_{i*}$)
the maximal probability of Alice (Bob) to bias the outcome to $i$.
In SCF the bias is defined as $\epsilon\hat{=}\max\left\{ \epsilon_{*0},\,\epsilon_{*1},\,\epsilon_{0*},\,\epsilon_{1*}\right\} $,
while in WCF the maximum is taken over only two of the biases: If,
for example, Alice prefers the outcome $0$, then the bias equals
$\max\left\{ \epsilon_{*0},\,\epsilon_{1*}\right\} $. A protocol is said to be fair
whenever both parties enjoy the same bias.

In classical settings, given unlimited computational power, a dishonest
party can always bias the outcome as it desires \cite{Kilian}. In
contrast, this is not the case in quantum settings. Aharonov \emph{et
al}. formulated the first (non-trivial) quantum SCF protocol in 2000
\cite{Aharonov}. This protocol, which achieves a bias of $0.354$ \cite{Spekkens -2},
began the quest for a SCF protocol with a vanishing bias. First, Spekkens
and Rudolph devised a protocol with a bias of $0.309$ \cite{Spekkens -2}.
Ambainis \cite{Ambainis} and independently Spekkens and Rudolph \cite{Spekkens -1}
soon afterwards introduced protocols pushing the bias to as low as
$1/4$. However, the prospects of further progress were soon shadowed
by two key results. Ambainis proved that any protocol with a bias
of $\epsilon$, whether strong or weak, must consist of at least $\Omega(\log\log\epsilon^{-1})$
rounds of communication \cite{Ambainis}, while Kitaev proved that
the bias of any quantum SCF protocol is bounded by $\left(\sqrt{2}-1\right)/2\simeq0.207$
\cite{Kitaev}. Until recently, it was not known whether this bound
can be saturated or whether the bias of $1/4$ is optimal. This point
has now been settled by Chailloux and Kerenidis who have presented
a protocol that saturates Kitaev's bound \cite{Kerenidis}, based
on Mochon's work proving the possibility of quantum WCF with arbitrarily
small bias \cite{Mochon}.

Quantum WCF was first analyzed by Spekkens and Rudolph in 2001, who
introduced a family of protocols that achieves a bias of $\left(\sqrt{2}-1\right)/2$
\cite{Spekkens}. (Previously Goldenberg \emph{et al}. analyzed the
problem of quantum gambling \cite{Gambling}, which is a closely related
cryptographic task.) This result was subsequently improved upon by
Mochon who considered WCF protocols with an infinite number of rounds
\cite{Mochon_-2,Mochon_-1}, eventually culminating in the aforementioned
result \cite{Mochon}. In addition, quantum SCF and WCF have also
been studied in the multi-party \cite{Multi-party} multi-outcome
scenario \cite{Barrett & Massar I,Barrett & Massar II} and most recently
in both \cite{Dice rolling,Ganz}.\\

Even though from a purely theoretical viewpoint a lot of progress
has been made in our understanding of quantum CF, most quantum CF
protocols are impractical due to the non-ideal conditions prevalent
in any real-life implementation. These include uncertainties in the
preparation and measurement of states, whether inherent or due to
noise, as well as noise and losses in the quantum channels and the
quantum memory storage. In the sending of quantum information over
long distances the most common source of malfunctions is losses. The
problem with losses is that they introduce a finite probability for
an indefinite outcome -- actually no outcome at all -- even when both
parties are honest, so that there is always a non-vanishing chance
for the protocol to be aborted (i.e. $P_{\perp}=1-P_{00}-P_{11}>0$).
As pointed out in \cite{Barrett & Massar I}, one way to avoid this
situation, is to restart the protocol each time an indefinite outcome
occurs, but this in turn affords a dishonest party very simple cheating
strategies, which may even go so far as to enable it to bias the outcome
to its choosing. Remarkably, Berl\'{i}n \emph{et al}. have recently devised
a `loss-tolerant' SCF protocol \cite{Berlin,Berlin experiment} (see
also \cite{Massar 2}). That is, a protocol whose bias remains unchanged
even if we allow for the protocol to be restarted. However, the loss-tolerance
came at a price: the protocol achieves a comparatively high bias of
$0.4$. It may well be that there is always a price to be paid. Specifically,
it could be that loss-tolerant SCF cannot saturate Kitaev's bound.
Indeed, at the end of their paper Berl\'{i}n \emph{et al}. raise the question
of whether it is possible to devise a loss-tolerant protocol with
a lower bias than theirs.

In this paper we answer this question in the affirmative by introducing
a family of loss-tolerant SCF protocols, which outperforms Berl\'{i}n
\emph{et al}.'s protocol. Each member in the family differs in the
number of qubits employed. In the one qubit case we achieve the same
bias as Berl\'{i}n \emph{et al.}, $\epsilon=0.4$, while for two qubits
the bias reduces to $0.3975$. Numerical evidence based on semi-definite
programming \cite{SDP}
suggests that the bias continues to decrease as the number of qubits
is increased, but at a rapidly decreasing rate. Our protocol bears
some similarity to the original BB84 CF protocol \cite{BB84} and
its various derivatives \cite{Aharonov,Ambainis,Berlin}, but significantly
differs in that it is not based on bit-commitment.

\section{A family of loss-tolerant protocols}

The protocols read as follows:
\begin{enumerate}
\item Alice selects $N$ orientations $\hat{\alpha}_{1}$ to $\hat{\alpha}_{N}$,
where each of the $\hat{\alpha}_{i}$ is (uniformly) randomly picked
from a set of four predetermined orientations $\hat{n}$, $-\hat{n}$,
$\hat{m}$, and -$\hat{m}$ (which are known to Bob). Alice prepares
$N$ qubits polarized along these orientations, i.e. the first qubit
is polarized along $\hat{\alpha}_{1}$, the second along $\hat{\alpha}_{2}$,
etc., and sends these $N$ qubits to Bob.
\item Bob selects $N$ orientations $\hat{\beta}_{1}$ to $\hat{\beta}_{N}$,
where each of the $\hat{\beta}_{i}$ is (uniformly) randomly picked
from the set of two orientations $\hat{n}$ and $\hat{m}$, and then
measures the polarization of the first qubit along $\hat{\beta}_{1}$,
the polarization of the second along $\beta_{2}$, etc.
\item If all the measurements are successful (i.e. he has detected the $N$
qubits and has obtained $N$ definite outcomes), he asks Alice to
proceed with the protocol, otherwise, he asks her to restart the
protocol (i.e. repeat step 1).
\item Alice sends Bob a randomly selected classical bit $c$.
\item Let $r_{1}$ to $r_{N}$ denote the outcomes of Bob's $N$ measurements.
The outcome of the coin flip $o$ is given by $o=c\oplus\bigl(\bigoplus_{i=1}^{N}r_{i}\bigr)$.
Bob informs Alice of his choice of orientations, $\hat{\beta}_{i}$, and the corresponding outcomes, $r_{i}$.
\item Alice aborts whenever there is at least one qubit that Bob claims
to have successfully measured for which $\left(1-2r\right)\hat{\beta}=-\hat{\alpha}$.
\end{enumerate}
The loss tolerance of the protocol comes into play at step 3, where
Bob asks Alice to restart the protocol whenever one or more of his measurements are unsuccessful, in which case the outcomes of the successful measurements are discarded. That is, Bob must successfully
measure $N$-qubits in a single run of the protocol.
Also note that we do not fix the angle between the axes $\hat{n}$
and $\hat{m}$. Indeed, this angle is a free a parameter. In particular,
it turns out that by manipulating it we can make the protocol `fair'
in the sense that Alice's and Bob's maximal biases are equal.

\begin{figure}
\center{ \includegraphics[scale=0.52]{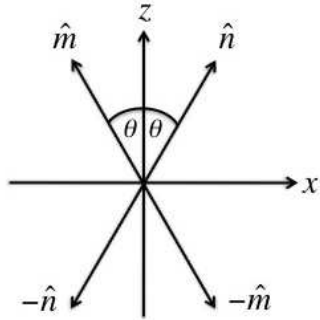}

\caption{Alice's preparation. Each qubit that Alice prepares is polarized along
one of the four axes $\pm\hat{n}$ and $\pm\hat{m}$.}

}
\end{figure}

\section{Alice's maximal bias}

It will prove convenient to choose the coordinate system such that
$\hat{n}$ and $\hat{m}$ lie on the $x\, z$ plane, spanning an angle
of $\theta$, $-\theta$, respectively, from the $z$ axis (see Fig. 1).\\

Since Bob is honest he will measure each qubit along one of the two
axes $\hat{n}$ and $\hat{m}$ with equal probability. Suppose that
Alice wishes to bias the outcome to 0. With no loss of generality
we assume that Alice selects $c=0$. Then the probability that she
is successful equals \begin{equation}
P_{*0}^{\left(N\right)}=\max_{\rho}\frac{1}{2^{N}}\sum_{\hat{b}_{1}=\hat{n},\,\hat{m}}\dots\sum_{\hat{b}_{n}=\hat{n},\,\hat{m}}P\Bigl(\bigoplus_{i=1}^{N}r_{i}=0\mid\bigl\{ \hat{b}_{1},\,\hat{b}_{2},\,\dots,\,\hat{b}_{N}\bigr\} ,\,\rho\Bigr)\,;\label{Alice's bias general form}\end{equation}
 the superscript $N$ serving to denote the number of qubits employed
in the protocol. Introducing the operator \begin{equation}
\Pi_{N}\hat{=}\frac{1}{2^{N}}\sum_{\hat{b}_{1}=\hat{n},\,\hat{m}}\sum_{s_{1}=\pm1}\dots\sum_{\hat{b}_{n}=\hat{n},\,\hat{m}}\sum_{s_{N}=\pm1}\Theta\left(s_{1}\cdot s_{2}\cdot\dots\cdot s_{N}\right)\bigotimes_{i=1}^{N}\bigl|\uparrow_{s_{i}\hat{b}_{i}}\bigr\rangle \bigl\langle \uparrow_{s_{i}\hat{b}_{i}}\bigr|\,,\label{Honest Bob's measurement}\end{equation}
 where $\Theta\left(x\right)$ is the Heaviside step function, we
have that \begin{equation}
P_{*0}^{\left(N\right)}=\max_{\rho}\mathrm{Tr}\left(\rho\Pi_{N}\right)\,.\label{Alice's bias general form II}\end{equation}
 Clearly, the maximum obtains when $\rho$ equals the (normalized)
eigenvector (or any one of the eigenvectors) of $\Pi_{N}$ corresponding
to the greatest eigenvalue. Making use of the fact that \begin{equation}
\left|\uparrow_{\pm\hat{n}}\right\rangle \left\langle \uparrow_{\pm\hat{n}}\right|+\left|\uparrow_{\pm\hat{m}}\right\rangle \left\langle \uparrow_{\pm\hat{m}}\right|=\mathds{1}\pm\cos\left(\theta\right)\sigma_{z}\label{Spin identities}\end{equation}
 (and $s_{1}\cdot s_{2}\cdot\dots\cdot s_{N}=1$ since Alice wishes
to bias the outcome to $0$), eq. (2) simplifies to \begin{eqnarray}
\Pi_{N} & = & \frac{1}{2^{N}}\sum_{s_{1}=\pm1}\dots\sum_{s_{N}=\pm1}\Theta\left(s_{1}\cdot s_{2}\cdot\dots\cdot s_{N}\right)\bigotimes_{i=1}^{N}\bigl(\mathbf{\mathds{1}}_{i}+s_{i}\cos\left(\theta\right)\sigma_{z}^{\left(i\right)}\bigr)\nonumber \\
 & = & \frac{1}{2^{N}}\sum_{s_{1}=\pm1}\dots\sum_{s_{N}=\pm1}\Theta\left(s_{1}\cdot s_{2}\cdot\dots\cdot s_{N}\right)\Bigl(\mathbf{\mathds{1}}+\sum_{i=1}^{N}s_{i}\cos\left(\theta\right)\Sigma_{z}^{\left(i\right)}+2\sum_{i=1}^{N}\sum_{j=i+1}^{N}s_{i}s_{j}\cos^{2}\left(\theta\right)\Sigma_{z}^{\left(i\right)}\Sigma_{z}^{\left(j\right)}\Bigr.\nonumber \\
 &  & \Bigl.+\dots+\cos^{N}\left(\theta\right)\prod_{i=1}^{N}\Sigma_{z}^{\left(i\right)}\Bigr)\nonumber \\
 & = & \frac{1}{2}\Bigl(\mathbf{\mathds{1}}+\cos^{N}\left(\theta\right)\bigotimes_{i=1}^{N}\sigma_{z}^{\left(i\right)}\Bigr)\,.\label{Honest Bob's measurement II}\end{eqnarray}
 Here we use the notation \begin{equation}
\Sigma_{\mathfrak{a}}^{\left(i\right)}\hat{=}\mathds{1}_{1}\otimes\dots\otimes\mathds{1}_{i-1}\otimes\sigma_{z}^{\left(i\right)}\otimes\mathds{1}_{i+1}\otimes\dots\otimes\mathds{1}_{N}\,,\qquad \mathfrak{a}=x,\, y,\, z\label{Sigma_a}\end{equation}
 with $\mathrm{\mathds{1}}_{i}$ denoting the identity operator on
the Hilbert space of the $i\,$th qubit. The eigenvalues of $\Pi_{N}$
equal $\left(1\pm\cos^{N}\left(\theta\right)\right)/2$. The resulting
biases are thus given by \begin{equation}
P_{*0}^{\left(N\right)}=P_{*1}^{\left(N\right)}=\frac{1}{2}\left(1+\cos^{N}\left(\theta\right)\right)\,,\label{Alice's bias}\end{equation}
 since the probability of biasing to $0$ and $1$ are patently equal.

\section{Bob's maximal bias}

In the following it will prove economical to employ the following
notation: $\bigl|\psi_{0}^{\left(0\right)}\bigr\rangle \hat{=}\bigl|\uparrow_{\hat{n}}\bigr\rangle $,
$\bigl|\psi_{0}^{\left(1\right)}\bigr\rangle \hat{=}\bigl|\downarrow_{\hat{n}}\bigr\rangle $,
$\bigl|\psi_{1}^{\left(0\right)}\bigr\rangle \hat{=}\bigl|\uparrow_{\hat{m}}\bigr\rangle $,
$\bigl|\psi_{1}^{\left(1\right)}\bigr\rangle \hat{=}\bigl|\downarrow_{\hat{m}}\bigr\rangle $,
so that the basis $\left|\uparrow_{\hat{n}}\right\rangle $, $\left|\downarrow_{\hat{n}}\right\rangle $
($\left|\uparrow_{\hat{m}}\right\rangle $, $\left|\downarrow_{\hat{m}}\right\rangle $)
is denoted by $0$ ($1$). In addition, we define $\bigl|\psi_{\mathbf{b}}^{\left(\mathbf{r}\right)}\bigr\rangle \hat{=}\bigotimes_{i=1}^{N}\bigl|\psi_{b_{i}}^{\left(r_{i}\right)}\bigr\rangle $,
where $\mathbf{b}\hat{=}\left(b_{1},\, b_{2},\,\dots,\, b_{N}\right)$
and $\mathbf{r}\hat{=}\left(r_{1},\, r_{2},\,\dots,\, r_{N}\right)$ are
binary $N$-tuples, i.e. $b_{i},\, r_{i}\in\{0,\,1\}$.\\

The loss-tolerant nature of the protocol allows (a dishonest) Bob
to carry out a measurement at step 2 to decide whether to keep the
$N$ the qubits. Only when he has decided to keep them does he proceed
to step 4. Then, depending on the value of the classical bit $c$ (received at step
4), he will carry out another measurement on the $N$ qubits at step
5. The outcome of this measurement instructs him what $N$-tuples $\mathbf{b}$
and $\mathbf{r}$ to tell Alice that he selected and (supposedly)
obtained, respectively. More specifically, at step 2 Bob will carry
out a two-outcome POVM with elements $\Pi_{\mathrm{p}}$, $\Pi_{\mathrm{rs}}=\mathds{1}-\Pi_{\mathrm{p}}$.
If he obtains the outcome associated with $\Pi_{\mathrm{rs}}$ he
asks Alice to restart the protocol. Otherwise, if he obtains the outcome
associated with $\Pi_{\mathrm{p}}$, he keeps the qubits and they
proceed to step 4. At step 5 Bob already knows the value of $c$.
Let us assume that he would like to bias the outcome to $0$, then
to optimize his chances of being successful he will have to tell Alice
announce an $N$-tuple $\mathbf{r}$ such that $\bigoplus_{i}r_{i}=0\oplus c=c$.
He will then carry out an additional POVM on the $N$-qubits with
$2^{2N-1}$ outcomes, which instructs him what $N$-tuples $\mathbf{b}$
and $\mathbf{r}$ to announce. We will denote this second POVM by
$\Pi_{0\, c}^{\mathbf{r}\,\mathbf{b}}$, where the subscripts $0$
and $c$ correspond to the value to which Bob wants to bias the coin
and the value of the classical bit sent by Alice, and the superscripts
$\mathbf{b}$ and $\mathbf{r}$ correspond to the choice of bases
and the associated outcomes that he sends Alice. Hence, a cheating
strategy designed to obtain the outcome $0$ consists
of a set of three POVMs with elements $\{ \Pi_{\mathrm{p}},\,\Pi_{\mathrm{rs}}\} $,
$\{ \Pi_{0\,0}^{\mathbf{r}\,\mathbf{b}}\mid\bigoplus r_{i}=0\} $
and $\{ \Pi_{0\,1}^{\mathbf{r}\,\mathbf{b}}\mid\bigoplus r_{i}=1\} $.

In the following it will facilitate matters
to introduce the positive operators $M_{0\, c}^{\mathbf{r}\,\mathbf{b}}\hat{=}\sqrt{\Pi_{\mathrm{p}}}\Pi_{0\, c}^{\mathbf{r}\,\mathbf{b}}\sqrt{\Pi_{\mathrm{p}}}$,
which we note satisfy \begin{equation}
\sum_{\left\{ \mathbf{b}\right\} }\sum_{\{ \mathbf{r}\mid\bigoplus_{i}r_{i}=c\} }M_{0\, c}^{\mathbf{r}\,\mathbf{b}}=\sqrt{\Pi_{\mathrm{p}}}\sum_{\left\{ \mathbf{b}\right\} }\sum_{\{ \mathbf{r}\mid\bigoplus_{i}r_{i}=c\} }\Pi_{0\, c}^{\mathbf{r}\,\mathbf{b}}\sqrt{\Pi_{\mathrm{p}}}=\Pi_{\mathrm{p}}\,.\label{POVM identity}\end{equation}

Suppose now that Alice prepared at step 1 the state $\bigl|\psi_{\mathbf{a}}^{\left(\mathbf{s}\right)}\bigr\rangle$,
and, having been asked to proceed with the protocol, sends Bob the
classical bit $c$ at step 4. Bob gets caught cheating when for one
or more of the qubits, $b_{i}=a_{i}$ and $r_{i}=s_{i}\oplus1$. Bob's
minimal probability of being caught cheating therefore equals \begin{equation}
1-P_{0*}^{\left(N\right)}=\frac{1}{2^{2N+1}}\min_{\left\{ M_{0\, c}^{\mathbf{r}\,\mathbf{b}}\right\} }\sum_{c=0,\,1}\sum_{\left\{ \mathbf{a}\right\} }\sum_{\left\{ \mathbf{s}\right\} }\sum_{\left\{ \mathbf{b}\right\} }\sum_{\{\mathbf{r}\mid\bigoplus_{i}r_{i}=c\}}\Theta\Bigl(\sum_{j=1}^{N}\delta_{b_{j},\, a_{j}}\cdot\delta_{r_{j},\, s_{j}\oplus1}\Bigr)\frac{\bigl\langle\psi_{\mathbf{a}}^{\left(\mathbf{s}\right)}\bigr|M_{0\, c}^{\mathbf{r}\,\mathbf{b}}\bigl|\psi_{\mathbf{a}}^{\left(\mathbf{s}\right)}\bigr\rangle}{\bigl\langle\psi_{\mathbf{a}}^{\left(\mathbf{s}\right)}\bigr|\Pi_{\mathrm{p}}|\psi_{\mathbf{a}}^{\left(\mathbf{s}\right)}\bigr\rangle}\,,\label{Bob's bias general form}\end{equation}
 where the summation is carried out over the set of all possible binary
$N$-tuples, $\{\mathbf{a}\}$, $\{\mathbf{b}\}$, $\{\mathbf{s}\}$,
and $\{\mathbf{r}\mid\bigoplus_{i}r_{i}=c\}$. The Heaviside step-function,
additionally defined such that $\Theta\left(0\right)\hat{=}0$, serves
to guarantee that only terms, which satisfy $b_{i}=a_{i}$ and $r_{i}=s_{i}\oplus1$
for at least one $i\in\{1,\,2,\,\dots,\, n\}$, contribute. Finally,
the $2^{2N+1}$ factor is just the number of possible choices for
the triplet $c$, $\mathbf{a}$, and $\mathbf{s}$.

Clearly, no value of $c$ is in any way preferable for Bob, nor is
any orientation or any particular qubit. This implies the existence
of an optimal symmetric cheating strategy in the sense that all of
the POVM elements (pertaining to both the POVM carried out when $c=0$
and the POVM carried out when $c=1$) contribute equally. To see this,
we first assume the existence of an optimal (possibly asymmetric)
strategy. Let $\{ \tilde{M}_{0\,0}^{\mathbf{r}\,\mathbf{b}}\} $
and $\{ \tilde{M}_{0\,1}^{\mathbf{r}\,\mathbf{b}}\} $
denote the corresponding two sets of positive operators. Then,
for any binary $N$-tuple $\mathbf{u}$, another optimal
cheating strategy is obtained by the transformation \begin{equation}
\tilde{M}_{0\, c}^{\mathbf{r}\,\mathbf{b}}\rightarrow M_{0\; c}^{\mathbf{r}\,\mathbf{b}\oplus\mathbf{u}}=\Sigma_{z}^{\mathbf{u}}\tilde{M}_{0\, c}^{\mathbf{r}\,\mathbf{b}}\Sigma_{z}^{\mathbf{u}}\,,\label{Rotation operators I}\end{equation}
 where $\Sigma_{\mathfrak{a}}^{\mathbf{u}}\hat{=}\Pi_{i=1}^{N}\left.\Sigma_{\mathfrak{a}}^{\left(i\right)}\right.^{u_{i}}$
($\mathfrak{a}=x,\, y,\, z$) and $\mathbf{a}\oplus\mathbf{b}\hat{=}\left(a_{1}\oplus b_{1},\,\dots,\, a_{N}\oplus b_{N}\right)$,
corresponding to rotations by $\pi$ about the $z$ axes of the coordinate
systems of the set of qubits $\left\{ i\mid u_{i}=1\right\} $. Similarly,
for any binary $N$-tuple $\mathbf{u}$, we obtain yet another
optimal cheating strategy via \begin{equation}
\tilde{M}_{0\, c}^{\mathbf{r}\,\mathbf{b}}\rightarrow M_{0\, c\oplus(\bigoplus_{i}u_{i})}^{\mathbf{r}\oplus\mathbf{u}\,\mathbf{b}\oplus\mathbf{u}}=\Sigma_{x}^{\mathbf{u}}\tilde{M}_{0\, c}^{\mathbf{r}\,\mathbf{b}}\Sigma_{x}^{\mathbf{u}}\,,\label{Rotation operators II}\end{equation}
 corresponding to rotations by $\pi$ about the $x$ axes of the coordinate
systems of the set of qubits $\left\{ i\mid u_{i}=1\right\} $. (When
$\bigoplus_{i}u_{i}=1$ we switch from a POVM corresponding to one
value of $c$ to a POVM corresponding to the other value.)

Now a strategy in which Bob chooses at random between different optimal
strategies is also optimal. By choosing uniformly at random between
optimal strategies related by the transformations eqs. (\ref{Rotation operators I})
and (\ref{Rotation operators II}), Bob obtains an optimal strategy
characterized by the positive operators \begin{equation}
M_{0\, c}^{\mathbf{r}\,\mathbf{b}}=\frac{1}{4^{N}}\sum_{\left\{ \mathbf{u}\right\} }\sum_{\left\{ \mathbf{w}\right\} }\Sigma_{x}^{\mathbf{u}}\Sigma_{z}^{\mathbf{w}}\tilde{M}_{0\; c\oplus(\bigoplus_{i}u_{i})}^{\mathbf{r}\oplus\mathbf{u}\;\mathbf{b}\oplus\mathbf{w}\oplus\mathbf{u}}\Sigma_{x}^{\mathbf{u}}\Sigma_{z}^{\mathbf{w}}\,;\label{Symmetric POVM}\end{equation}
the only subtle point in the above argument concerns those
transformations given by eq. (\ref{Rotation operators II}) that modify
the value $c$ and exchange between elements in $\{ \tilde{M}_{0\,0}^{\mathbf{r}\,\mathbf{b}}\} $
and $\{ \tilde{M}_{0\,1}^{\mathbf{r}\,\mathbf{b}}\} $.
Nevertheless, this does not pose a problem since in an optimal cheating
strategy the overall contribution to the cheating probability when
$c=0$ and $c=1$ must be equal, and, moreover, eqs. (\ref{Rotation operators II})
and the invariance of $\Pi_{\mathrm{p}}$ under the application of the rotation operators,
imply that in an optimal cheating strategy the sets $\{ \tilde{M}_{0\,0}^{\mathbf{r}\,\mathbf{b}}\} $
and $\{ \tilde{M}_{0\,1}^{\mathbf{r}\,\mathbf{b}}\} $
can be obtained from one another via the transformation eq. (\ref{Rotation operators II}).
Finally, we note that this pair of sets, eq. (\ref{Symmetric POVM}),
can be characterized by any of the positive operators within the sets,
say $M_{0\,0}^{\mathbf{0}\,\mathbf{0}}$
($\mathbf{0}\hat{=}\left(0,\,\dots,\,0\right)$). All other positive
operators (including those corresponding to $c=1$) can be obtained
from it by the transformations eqs. (\ref{Rotation operators I})
and (\ref{Rotation operators II}). In appendix A we prove that eqs.
(\ref{POVM identity}), (\ref{Rotation operators I}), (\ref{Rotation operators II})
and (\ref{Symmetric POVM}) imply that one can take $\Pi_{\mathrm{p}}=\mathds{1}$.
This means that Bob stands nothing to gain by performing a measurement
on the qubits prior to receiving the value of the classical bit $c$.\\

The problem of optimizing Bob's bias can be cast as an SDP (see \cite{SDP} for an introduction). Using the fact that we
can set $\Pi_{\mathrm{p}}=\mathds{1}$ (and recalling that the rotation operators switch between all of Alices' preparations), the right-hand side of eq. (\ref{Bob's bias general form})
can be reexpressed as $\mathrm{Tr}\left(M_{0\,0}^{\mathbf{0}\,\mathbf{0}}\Lambda_{N}\left(\theta\right)\right)$,
with \begin{equation}
\Lambda_{N}\left(\theta\right)=\frac{1}{2^{2N+1}}\sum_{c=0,\,1}\sum_{\left\{ \mathbf{a}\right\} }\sum_{\left\{ \mathbf{s}\right\} }\sum_{\left\{ \mathbf{b}\right\} }\sum_{\{ \mathbf{r}\mid\bigoplus_{i}r_{i}=c\} }\Theta\Bigl(\sum_{j=1}^{N}\delta_{b_{j},\, a_{j}}\cdot\delta_{r_{j},\, s_{j}\oplus1}\Bigr)\bigl|\psi_{\mathbf{a}\oplus\mathbf{b}}^{\left(\mathbf{s}\oplus\mathbf{r}\right)}\bigr\rangle \bigl\langle \psi_{\mathbf{a}\oplus\mathbf{b}}^{\left(\mathbf{s}\oplus\mathbf{r}\right)}\bigr|\,.\label{Lambda}\end{equation}
 The SDP then reads

\begin{eqnarray}
 & P_{0*}^{\left(N\right)}=\max_{M_{0\,0}^{\mathbf{0}\,\mathbf{0}}}\left(1-\mathrm{Tr}\left(M_{0\,0}^{\mathbf{0}\,\mathbf{0}}\Lambda_{N}\left(\theta\right)\right)\right)\label{General dual problem}\\
 & \mathrm{subject\; to}\qquad2^{N-1}\mathrm{Tr}\left(M_{0\,0}^{\mathbf{0}\,\mathbf{0}}\right)=1,\quad\mathrm{Tr}\bigl(M_{0\,0}^{\mathbf{0}\,\mathbf{0}}\bigotimes_{i=1}^{N}\sigma_{z}^{\left(i\right)}\bigr)=0,\quad M_{0\,0}^{\mathbf{0}\,\mathbf{0}}\geq 0\,.\nonumber \end{eqnarray}
 The derivation of the first two constraints is given in Appendix A.

Now problems of this type,  have associated
with them a dual problem. The solution of this dual problem bounds
from above the solution of the of the original problem, \cite{SDP}, which we shall refer to as the `primal' problem. It is given by \begin{eqnarray}
 & \min_{\lambda_{i}}\left(1-\frac{1}{2}\lambda_{1}\right)\label{General primal problem}\\
 & \mathrm{subject\; to}\qquad\Lambda_{N}\left(\theta\right)-\lambda_{1}\mathds{1}+\lambda_{2}\bigotimes_{i=1}^{N}\sigma_{z}^{\left(i\right)}\geq0\,,\nonumber \end{eqnarray}
 where the variables of the dual problem, the $\lambda_{i}$, are
real scalars.

\subsection{The single qubit case}

It is straightforward to solve both eqs. (\ref{General dual problem}) and (\ref{General primal problem}) in the single qubit
case. The solution is given by\begin{equation}
P_{0*}^{\left(1\right)}=P_{1*}^{\left(1\right)}=\frac{1}{4}\left(3+\sin\left(\theta\right)\right)\,,\label{Bob's bias single qubit}\end{equation}
 where the second equality follows from the equality of the probabilities
of biasing to $0$ and $1$, and is obtained for $2M_{0\,0}^{0\,0}=\mathds{1}+\sigma_{x}$.
Hence, Bob's strategy consists of measuring the polarization of the
qubit along the $x$ axis.

\subsection{The two-qubit case}

In the two qubit case Bob measures an eight outcome POVM $M_{0\quad c}^{\left(r_{1},\, r_{1}\oplus c\right)\,\mathbf{b}}$
(recall that we have assumed that Bob wants to bias the outcome to
$0$). By introducing a new set of Lagrange multipliers $\xi=\lambda_{1}-\lambda_{2}$
and $\chi=\lambda_{1}+\lambda_{2}$, the dual problem can be reexpressed
as \begin{eqnarray}
 & \min_{\xi,\,\chi}\left(1-\frac{1}{4}\left(\xi+\chi\right)\right)\label{Dual problem two qubits II}\\
 & \mathrm{subject\; to}\qquad\Lambda_{2}\left(\theta\right)-\xi\left(\left|\uparrow\uparrow\right\rangle \left\langle \uparrow\uparrow\right|+\left|\downarrow\downarrow\right\rangle \left\langle \downarrow\downarrow\right|\right)-\chi\left(\left|\uparrow\downarrow\right\rangle \left\langle \uparrow\downarrow\right|+\left|\downarrow\uparrow\right\rangle \left\langle \downarrow\uparrow\right|\right)\geq0\,.\nonumber \end{eqnarray}
 The solution obtains when \begin{equation}
\det\left(\Lambda_{2}\left(\theta\right)-\xi\left(\left|\uparrow\uparrow\right\rangle \left\langle \uparrow\uparrow\right|+\left|\downarrow\downarrow\right\rangle \left\langle \downarrow\downarrow\right|\right)-\chi\left(\left|\uparrow\downarrow\right\rangle \left\langle \uparrow\downarrow\right|+\left|\downarrow\uparrow\right\rangle \left\langle \downarrow\uparrow\right|\right)\right)=0\,,\label{Determinant equation}\end{equation}
 i.e. when the lowest eigenvalue of the constraint matrix eq. (\ref{Dual problem two qubits II})
equals zero. Solving for $\chi$ in terms of $\xi$ we get \begin{equation}
\chi=\frac{\left(\cos\left(2\theta\right)+7\right)\xi^{2}-3\left(\cos\left(2\theta\right)+3\right)\xi}{8\xi^{2}+\left(\cos\left(2\theta\right)-13\right)\xi-3\left(\cos\left(2\theta\right)-1\right)}\,.\label{Solution of determinant equation}\end{equation}
 (There is another solution $\chi=1$, but it can be shown that in
this case the constraint matrix always admits a negative eigenvalue.)
Plugging back into eq. (\ref{Dual problem two qubits II}), taking
the derivative with respect to $\xi$, and equating to zero, we get
a fourth order equation in $\xi$ \begin{equation}
64\xi^{4}+16\left(\cos\left(2\theta\right)-13\right)\xi^{3}+\left(\cos\left(4\theta\right)-56\cos\left(2\theta\right)+199\right)\xi^{2}-6\left(\cos\left(4\theta\right)-8\cos\left(2\theta\right)+7\right)\xi+9\left(\cos\left(4\theta\right)-1\right)=0\,.\label{Fourth order equation}\end{equation}
 When plugged back into eq. (\ref{Dual problem two qubits II}) three
of the four roots do not give rise to expressions smaller to Bob's
maximal bias in the single qubit case. See Fig. 2. Hence, none of
these three represents a solution of the primal problem since its
solution must bound from above the solution of the dual problem and
clearly Bob can always achieve a bias equal to that of the single
qubit case by simply not following the directions of the protocol
in the handling of only one of the qubits. It is straightforward to
show that the remaining eigenvalue satisfies the constraints of the
dual problem (i.e. all other three eigenvalues are positive), and
therefore gives rise to an upper bound on Bob's maximal bias.

\begin{figure}
\center{ \includegraphics{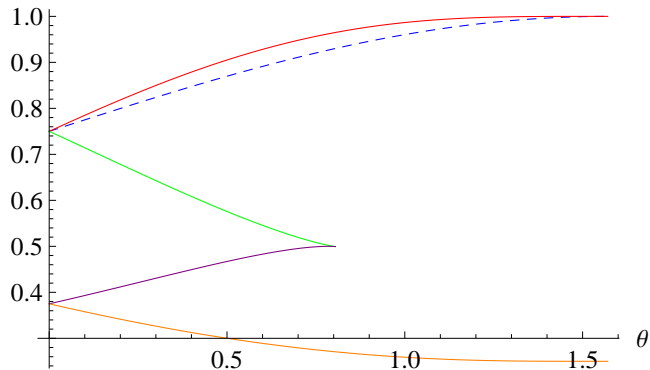}

\caption{The dashed curve depicts Bob's maximal probability of biasing the
outcome in the single qubit case, while each of the four other curves
depict eq. (\ref{Dual problem two qubits II}) as a function of a
different root of eq. (\ref{Fourth order equation}). Note that two
of the roots become complex beyond $\theta \simeq 0.8056$.}

}
\end{figure}

\section{Biases in the fair scenario}

To make the protocol fair, i.e. $P_{F}^{\left(N\right)}=P_{*i}^{\left(N\right)}=P_{j*}^{\left(N\right)}$,
we have the freedom to manipulate $\theta$. In this way, for a single
qubit we obtain $P_{*i}^{\left(1\right)}=P_{j*}^{\left(1\right)}=0.9$
($\theta\simeq36.87^{\circ}$). In the two qubit case, the solution
of the dual problem and Alice's maximal bias intersect for $\theta\simeq26.92^{\circ}$,
$\xi\simeq-0.2098$, $\chi\simeq0.6197$, $1-\left(\xi+\chi\right)/4\simeq$
$P_{*0}^{\left(2\right)}\simeq0.8975$. It remains to prove that this
intersection indeed corresponds to Bob's maximal bias, or, what is
the same thing, to show that for this value of the angle the solution
of the dual problem coincides with that of the primal problem.

To do so we conjecture that the solution of the primal problem, eq.
(\ref{General primal problem}) in the case $N=2$, is of the form
\begin{equation}
2M_{0\,0}^{\mathbf{0}\,\mathbf{0}}=\left|\upsilon\left(\theta\right)\right\rangle \left\langle \upsilon\left(\theta\right)\right|\,,\label{Solution of primal problem two qubits}\end{equation}
 where \begin{equation}
\left|\upsilon\left(\theta\right)\right\rangle =\frac{1}{\sqrt{2}}\cos\left(f\left(\theta\right)\right)\left|\uparrow\uparrow\right\rangle +\frac{1}{2}\left|\uparrow\downarrow\right\rangle +\frac{1}{2}\left|\downarrow\uparrow\right\rangle +\frac{1}{\sqrt{2}}\sin\left(f\left(\theta\right)\right)\left|\downarrow\downarrow\right\rangle \label{Solution of primal problem two qubits II}\end{equation}
 and $f$ is some real function of $\theta$. It is easy to verify
that eq. (\ref{Solution of primal problem two qubits}) satisfies
the constraints eq. (\ref{General primal problem}). As a functional
of $f$, the probability of biasing the outcome to $0$ (or $1$)
then reads  \begin{eqnarray}
P_{0*}^{\left(2\right)} & = & \frac{1}{64}\left(12\cos\left(2f\left(\theta\right)\right)\cos\left(\theta\right)+2\sin\left(2f\left(\theta\right)\right)\sin^{2}\left(\theta\right)+\cos\left(f\left(\theta\right)\right)\bigl(12\sqrt{2}\sin\left(\theta\right)+\sin\left(2\theta\right)\big)\right.\nonumber \\
 &  & \left.+\sin\left(f\left(\theta\right)\right)\bigl(12\sqrt{2}\sin\left(\theta\right)-\sin\left(2\theta\right)\bigr)-\cos\left(2\theta\right)+37\right)\,.\label{General form of Bob's bias two qubits}\end{eqnarray}
 For $\theta\simeq26.92^{\circ}$ $P_{0*}^{\left(2\right)}$ is maximized
for $f\simeq0.1177$ equaling $0.8975$ as anticipated.\\

We see that while Alice's maximal bias decreases with the increase
in the number of qubits, Bob's maximal bias increases as we go from
one to two qubits (see Fig. 2). For a greater number of qubits, numerical
based SDP evidence indicates the bias in the fair scenario continue
to decrease with the increase in the number of qubits (Bob's bias
increases), but at an increasingly slower rate. We were able to carry
out numerics for up to six qubits, and obtained $P_{F}^{\left(3\right)}\simeq0.8967$,
$P_{F}^{\left(4\right)}=0.8962$, $P_{F}^{\left(5\right)}\simeq89.60$,
$\left.P_{F}^{\left(6\right)}\right.\simeq0.8958$ (in the later case
$\theta\simeq15.89^{\circ}$).

\begin{figure}
\center{ \includegraphics{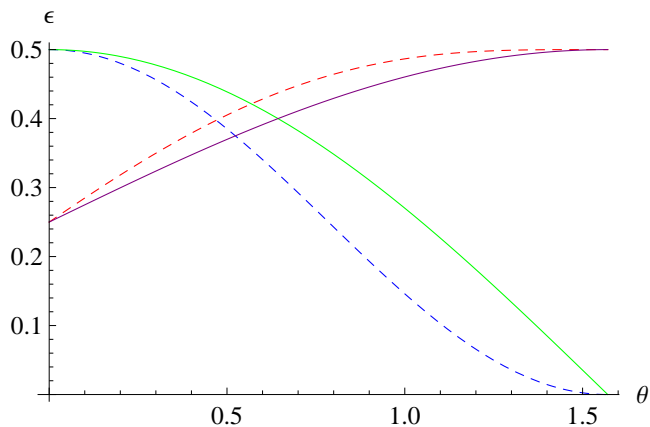}

\caption{Maximal biases as a function of $\theta$ for $N\leq2$ qubits. The
dashed (solid) decaying curve depicts Alice's bias for $N=2$ ($N=1$)
and the dashed (solid) rising curve depicts Bob's bias for $N=2$
($N=1$).}

}
\end{figure}

\section{Conclusion}

It is possible to overcome the problem of losses in quantum CF. A
loss-tolerant CF protocol has the property that its bias remains unchanged
even if we allow for it to be restarted whenever losses occur. However,
this robustness seems to come at a price. Berl\'{i}n \emph{et al}.'s loss-tolerant
SCF protocol achieves a relatively high bias of $0.4$.
In this paper, by presenting a family of loss-tolerant SCF protocols, we were able to show that Berl\'{i}n \emph{et al}.'s
result can be improved upon. Utilizing a single qubit we reproduced
their result, while utilizing a pair of qubits we obtained a bias
of $0.3975$. SDP based numerical evidence indicates that the bias
continues to decrease as the number of qubits is increased, but at
a rapidly decreasing rate.

In future work it should be interesting to determine the theoretical
limits on loss-tolerant CF protocols. Specifically, can Kitaev's bound
be saturated by a loss-tolerant SCF protocol? If not, what is the
bound on loss-tolerant SCF protocols? Furthermore, is it possible
to introduce a loss-tolerant WCF protocol? Two main difficulties
are apparent. First, at the end of a WCF protocol the losing party
usually verifies the outcome by measuring a quantum system that has
been kept in a quantum memory storage. Hence, in this scenario the
losing party can always avoid losing by claiming to have lost the
stored system. Second, the number of rounds of communication required
to realize a CF protocol with a bias of $\epsilon$ is of the order
of $\Omega(\log\log\epsilon^{-1})$. In particular, to achieve a loss-tolerant
WCF protocol with an arbitrarily small bias will require the protocol
to be impervious to losses occurring at any round, implying that a
dishonest party must not be capable of (probabilistically) inferring
whether it is going to win or lose at at any round before the last.
\begin{acknowledgments}
We wish to thank Stefano Pironio for helpful discussions and acknowledge
the support of the European Commission under the Integrated Project
Qubit Applications (QAP), funded by the IST Directorate (Contract
no. 015848). In addition, N. Aharon also acknowledges the support
of the Binational Science Foundation and The Wolfson Family Charitable Trust (Grant number 32/08). 
S. Massar and J. Silman also acknowledge
the support of the Inter-University Attraction Poles Programme (Belgian
Science Policy) under Project IAP-P6/10 (Photonics@be).
\end{acknowledgments}
\appendix

\section{}
Here we prove that $\Pi_{\mathrm{p}}$ can be set equal to the identity (see
remark below eq. (\ref{POVM identity})), and we prove the constraints
on primal problem, eq. (\ref{General dual problem}).

Recall that $ $\begin{equation}
\sum_{\left\{ \mathbf{b}\right\} }\sum_{\{ \mathbf{r}\mid\bigoplus_{i}r_{i}=c\} }M_{0\, c}^{\mathbf{r}\,\mathbf{b}}=\Pi_{\mathrm{p}}\,.\label{Sum over POVM elements}\end{equation}
 We would like to show that upon summation all Pauli basis vectors
composed of or one or more of the $\sigma_{x}^{\left(i\right)}$ or
$\sigma_{y}^{\left(i\right)}$ vanish. To see this, we note for every
element in the sum $M_{0\, c}^{\mathbf{r}\,\mathbf{b}}$ there
is another element $M_{0\, c}^{\mathbf{r}\,\mathbf{b}'}$, such that
$b_{j}'=b_{j}\oplus\delta_{ij}$. Algebraically, this second element
is identical to the first except that in its Pauli basis expansion
the coefficient of every basis vector composed of either $\sigma_{x}^{\left(i\right)}$
or $\sigma_{y}^{\left(i\right)}$ has the opposite sign, so that upon
summation they cancel each other. See eqs. (\ref{Rotation operators I})
and (\ref{Rotation operators II}).

It remains to show that the sum of all Pauli basis vectors composed
solely of one or more of the $\sigma_{z}^{\left(i\right)}$ and identity
operators vanish. We note that for every element in the sum $M_{0\, c}^{\mathbf{r}\,\mathbf{b}}$
there is another element $M_{0\, c}^{\mathbf{r}'\,\mathbf{b}}$, such
that $r_{k}'=r_{k}\oplus\delta_{ik}\oplus\delta_{jk}$ with $i\neq j$.
This second element is identical to the first except that in
its Pauli basis expansion the coefficients of basis vectors composed
of either $\sigma_{z}^{\left(i\right)}$ or $\sigma_{z}^{\left(j\right)}$
(and identity operators), but not both, have opposite signs. See eqs.
(\ref{Rotation operators I}) and (\ref{Rotation operators II}).
The vanishing of basis vectors composed of both $\sigma_{z}^{\left(i\right)}$
and $\sigma_{z}^{\left(j\right)}$ (and identity operators), but not
$\bigotimes_{i=1}^{N}\sigma_{z}^{\left(i\right)}$, then follows from
repeating this argument for all possible pairs of indices $k$ and
$l\neq k$. Hence, upon summation all Pauli basis vectors composed
of one or more of the $\sigma_{z}^{\left(i\right)}$, except $\bigotimes_{i=1}^{N}\sigma_{z}^{\left(i\right)}$,
vanish, and it follows that \begin{equation}
\sum_{\left\{ \mathbf{b}\right\} }\sum_{\{ \mathbf{r}\mid\bigoplus_{i}r_{i}=c\} }M_{0\, c}^{\mathbf{r}\,\mathbf{b}}=\alpha\mathds{1}+\gamma\otimes_{i=1}^{N}\sigma_{z}^{\left(i\right)}\,.\label{A2}\end{equation}
 However, from eq. (\ref{Rotation operators II}) we have that \begin{equation}
\sum_{\left\{ \mathbf{b}\right\} }\sum_{\{ \mathbf{r}\mid\bigoplus_{i}r_{i}=c\oplus 1\} }M_{0\, c\oplus1}^{\mathbf{r}\,\mathbf{b}}=\alpha\mathds{1}-\gamma\otimes_{i=1}^{N}\sigma_{z}^{\left(i\right)}\,.\label{A3}\end{equation}
 Since both eqs. (\ref{A2}) and (\ref{A3}) must be equal, the coefficient
of $\otimes_{i=1}^{N}\sigma_{z}^{\left(i\right)}$ must vanish and
$\Pi_{\mathrm{p}}$ is seen to be proportional to the identity. This implies
that Bob learns nothing from his first POVM with elements $\Pi_{\mathrm{p}}$
and $\Pi_{\mathrm{rs}}$, since both elements are proportional to
the identity. So that without loss of generality we can take $\Pi_{\mathrm{p}}$
equal to the identity.

Finally, since the sum is composed of $2^{2N-1}$
$2^{N}\times2^{N}$-dimensional matrices, and since in its Pauli basis
expansion each matrix admits the same coefficient for the identity,
it follows that $2^{N-1}\mathrm{Tr}\left(M_{0\, c}^{\mathbf{r}\,\mathbf{b}}\right)=1$.
Together, these last remarks imply the constraints in the SDP eq.
(\ref{General dual problem}).

\end{document}